# Lower Bounds for Zero Knowledge on the Internet

Joe Kilian [*]   Erez Petrank [†]   Charles Rackoff[‡]

June 2, 2018

**Abstract**

We consider zero knowledge interactive proofs in a richer, more realistic communication environment. In this setting, one may simultaneously engage in many interactive proofs, and these proofs may take place in an asynchronous fashion. It is known that zero-knowledge is not necessarily preserved in such an environment; we show that for a large class of protocols, it *cannot* be preserved. Any 4 round (computational) zero-knowledge interactive proof (or argument) for a non-trivial language $L$ is not black-box simulatable in the asynchronous setting.

## 1 Introduction

Zero knowledge [20] turned out to be a useful tool for many cryptographic applications. Many works have studied the numerous uses of zero knowledge proofs, and many other works have suggested how to improve the efficiency of these proofs. However, most of these works considered only the case where the proof stands alone, disconnected from the computing environment. An interesting question, which naturally arises these days, is how robust the notion of zero knowledge is in a broader setting. In particular, many computers today are connected through networks (from small local area networks to the entire Internet) in which connections are maintained in parallel asynchronous sessions. It is common to find several connections (such as FTP, Telnet, An internet browser, etc.) running together on a single workstation. Can zero knowledge protocols be trusted in such an environment?

The robustness of zero knowledge has been studied before in the "simple" case of parallel repetitions. It is often desirable to run a probabilistic protocol many times in parallel, usually in order to reduce the expected error of a single run. The alternative of running these protocols sequentially has a cost of increasing the number of rounds and is considered inefficient. It had been noted by several researchers that even in the parallel repetitions case the zero knowledge property does not necessarily hold. Goldreich and Krawczyk [16] proved a general lower bound: Any language that has a three round (black-box) zero-knowledge interactive proof with a small error probability (as can be obtained by parallel repetitions) is in BPP. Thus, for example, unless

---

[*]Yianilos Labs, `joe@pnylab.com`. Work done in part while at the NEC Research Institute.

[†]Dept. of Computer Science, Technion - Israel Institute of Technology, Haifa 32000, Israel. Email: erez@cs.technion.ac.il.

[‡]Dept. of Computer Science, University of Toronto, Toronto, Ontario, Canada M5S 3G4. Email: rackoff@cs.toronto.edu.

Graph Isomorphism is in BPP, the protocol of [17] for Graph Isomorphism does not remain zero knowledge when run many times in parallel. Several papers have dealt with this problem, usually by letting the verifier commit on it's (non-adaptive) questions in advance [1, 3, 15, 12].

Our initial feeling was that the protocols that keep their zero knowledge property when run in parallel should also remain zero knowledge even in a multi-session asynchronous environment. However, in this paper, we give some indications that this is not always the case.

Let us say a few words about what we need to show. In order to show that a protocol is zero knowledge in a modern networking environment, one must provide a proof that even within this complicated environment the protocol is still zero knowledge. But for us matters are simpler. We don't need to cover all facets of a networking environment. We only have to show that zero knowledge may fail in a specific setting that is part of this environment. Namely, the environment may be more hostile to the protocol than the specific case we study, but since a protocol fails even with a benign setting, it is definitely not zero knowledge when we extend the power of the environment. In particular, a networking environment may have various protocols running, multiple sessions, more than two parties involved, asynchronous setting, etc.

We show that zero knowledge may fail for a large class of protocols, even if we only run a single protocol between only two parties. We exploit the asynchronicity and the existence of multiple sessions. Clearly, if we also add other features, such as additional parties and other protocols running in parallel, then the security problems can only be amplified.

We show that any four-round black-box zero-knowledge proof with perfect completeness is not zero knowledge even in a very benign setting: The setting in which the protocol is run many times in an asynchronous environment, and an adversary (or the verifier) gets to choose which message gets to be delivered next. Actually, the setting is even more benign: We set a specific schedule, known in advance to both prover and verifier. Still, if the protocol can be (black-box) simulated, then the language must be in BPP.

**Theorem 1** *Suppose that $(P, V)$ is a 4-message interactive proof or argument for $L$ with perfect completeness and error at most $\frac{1}{2}$ and suppose that $(P, V)$ can be black-box simulated in the asynchronous setting. Then $L \in \mathcal{BPP}$.*

The $\frac{1}{2}$ in the above statement may be replaced by any constant less than 1, with essentially no change to our proofs; with slight care, any error bounded away from 1 by a non-negligible factor may be accommodated.

The above result holds for all extensions of concurrent zero-knowledge. In particular, it holds for *resettable zero-knowledge* [5].

Note that we get a separation between protocols that remain zero knowledge even under parallel repetition and protocols that remain zero knowledge in an asynchronous setting. Assuming the existence of one-to-one one-way functions, there exist 4-message (computational) zero knowledge arguments for all NP languages [12]. However, there are no 4-message zero knowledge arguments for languages outside of $\mathcal{BPP}$ that are black box simulatable in the asynchronous setting.

Some words on the terminology we are using. By zero knowledge we mean *computational* zero knowledge, i.e., the distribution output by the simulation is polynomial-time indistinguishable from the distribution of the views of the verifier in the original interaction. (Of-course, as the result is a lower bound, it holds also for perfect and statistical zero knowledge.) The prover may be infinitely powerful (i.e., an interactive proof) or it may be computationally bounded (i.e., an argument). We consider black box zero knowledge as defined by Goldreich and Oren [24, 18], and refined in [16].

## 1.1 Related work

Dwork, Naor, and Sahai [10] were the first to explore zero-knowledge in the asynchronous setting. They denoted zero-knowledge protocols that are robust to asynchronous composition *concurrent zero-knowledge* protocols. It was noticed in [10] that several known zero-knowledge proofs, with a straightforward adaptation of their original simulation to the asynchronous environment, may cause the simulator to work exponential time. Thus, it seems that the zero-knowledge property does not necessarily carry over to the asynchronous setting. In order to provide a protocol that may be used in a modern environment, they presented a compromise: a protocol that is not zero-knowledge in a fully asynchronous setting, but is zero-knowledge in an environment with bounds on the asynchronisity. In particular, they present a 4 round zero-knowledge argument for NP assuming that there are two constants $\alpha$ and $\beta$ such that the fastest message arrives within time at least $\alpha$ and the slowest message arrives within time at most $\beta$.

Dwork and Sahai [11] reduced the limitation on the asynchronisity. They presented a proof which has a preprocessing phase and a proof-body phase. Their proof is concurrent zero-knowledge argument for NP such that the $(\alpha, \beta)$ limitation is required only during the preprocessing stage. Then, the body of the proofs can be run in a fully asynchronous environment.

Our result is complementary to theirs, illustrating why it is difficult to achieve zero-knowledge in the asynchronous setting without using such an augmented model.

Several subsequent works have already appeared since the first publication of this paper [23]. One question that was raised was: does there exist a fully asynchronous (concurrent) zero-knowledge proof for NP?

The answer was given by Richardson and Kilian [25] who presented concurrent zero-knowledge arguments for all languages in NP, that is robust in the fully asynchronous setting. However, this protocol is not practical. It requires a polynomial number in $k$ of rounds, which makes it unacceptable in practice. $k$ here is the security parameter: the length of the input and the number of proofs that may run concurrently are bounded by a polynomial in $k$.

The upper bound on the number of rounds has been drastically improved by Kilian and Petrank [22]. It turns out that the number of rounds that suffices for constructing concurrent zero knowledge is any $m$ satisfying $m = \omega(\log^2 k)$.

Rosen [26] has improved the result described in this paper, i.e., the lower bound on the number of rounds required for concurrent zero-knowledge from 5 to 8. Canetti, Kilian, Petrank and Rosen [6] have substantially improved the lower bound to $\Omega(\log^k / \log \log k)$. The parameter $k$ is the security parameter. A polynomial in $k$ bounds the length of the inputs, the number of proofs that may start concurrently, and the time complexity that the parties spend in the protocol.

Other researchers have concentrated on presenting efficient concurrent zero-knowledge protocols for NP with weaker compromises on the asynchronisity of the environment. Crescenzo and Ostrovsky [8] presented a concurrent zero-knowledge argument for NP with a preprocessing phase. They removed the $(\alpha, \beta)$ constraint of [11], and the only requirement is that there must be a separating point in time between all preprocessing of all concurrent proofs, and all bodies of all proofs. Namely, the first body of any proof may start only after all preprocessing phases in all the proofs have completed. Dåmgard [9] and Canetti et. al. [5] have further reduced the limits on asynchronisity. They require that prior to the beginning of the proofs, all verifiers have deposited a public key in a public database. In [9] it is required that the public key is valid, i.e., that the verifier must know the secret key associated with it. In [5] this requirement is relaxed. The verifier only has to have a deposited string in the public database. Thus, these proofs are efficient, and make only the following compromise over full asynchronisity: all verifiers must be previously registered before any of them can engage in a proof. A verifier that has not been registered cannot join until all proofs are completed and then it must register itself before any new proof begins.

As an extension to concurrent zero-knowledge, Canetti et. al. presented *resettable zero-knowledge proofs* [5]. These are zero-knowledge proofs that on top of being concurrent, they remain zero-knowledge also when the verifier is allowed to run the prover repeatedly on a fixed (yet, randomly chosen) random tape. Our lower bound applies, of-course, also to resettable zero-knowledge.

Feige and Shamir have suggested to give up achieving full zero-knowledge in the asynchronous setting and presented a property of proofs called witness indistinguishability. They showed that witness indistinguishability is preserved also in the asynchronous setting [13].

The basic framework of the proof in this paper uses the ideas developed by Goldreich and Krawczyk [16]. Following their technique, we use a good simulator $S$ of an interactive proof (or argument) $(P, V)$ for a language $L$ to create an efficient prover $P_S$ that causes $V$ to accept reasonably often on inputs in $L$.

## 1.2 Guide to the paper

In Section 2 we discuss black-box simulatability and the framework used by Goldreich and Krawczyk. In Section 3 we show how to convert an asynchronous simulator into an efficient prover. In Sections 4 and 5 we analyze the success probability of this prover.

## 2 Preliminaries

### 2.1 Black-Box Zero-Knowledge

The initial definition of zero-knowledge [19] required that for any probabilistic polynomial time verifier $\hat{V}$, a simulator $S_{\hat{V}}$ exists that could simulate $\hat{V}$'s view. Goldreich and Oren [24, 18] propose a seemingly stronger, "better behaved" notion of zero-knowledge, known as *black-box* zero-knowledge. The basic idea behind black box zero-knowledge is that instead of having a new simulator $S_{\hat{V}}$ for each possible verifier, we have a single probabilistic polynomial time simulator $S$ that interacts with each possible $\hat{V}$. Furthermore, $S$ is not allowed to examine the internals of $\hat{V}$, but must simply look at $\hat{V}$'s input/output behavior. That is, it can have conversations with $\hat{V}$ and use these conversations to generate a simulation of $\hat{V}$'s view that is computationally indistinguishable from $\hat{V}$'s view of its interaction with $P$.

More formally, Goldreich and Krawczyk give the following version of this definition (notation changed for compatibility with our own), which avoids certain trivial problems in the original expositions.

**Definition 1** *([16], following [24, 18]) An interactive proof $\langle P, V \rangle$ is called* black-box simulation zero-knowledge *if for every polynomial $p$, there exists a probabilistic expected polynomial time oracle machine $S_p$ such that for any polynomial size verifier $\hat{V}$ that uses at most $p(n)$ random coins on inputs of length $n$, and for $x \in L$, the distributions $\langle P, \hat{V} \rangle(x)$ and $M_p^{\hat{V}}(x)$ are polynomially indistinguishable.*

**Remarks:** In the above definition, $\hat{V}$ may be thought of as a circuit (pedantically, a circuit family) that has access to random bits. The size of $\hat{V}$ refers to $\hat{V}$'s circuit size. It follows that $\hat{V}$ runs in polynomial time, which is necessary for a meaningful notion of computational zero knowledge, though not for statistical and perfect zero knowledge. By polynomially indistinguishable, we informally mean that no computationally bounded distinguisher can, for some constant $c$ and infinitely many $x \in L$, correctly guess whether a random sample came from $\langle P, \hat{V} \rangle(x)$ or $S_p^{\hat{V}}(x)$ with probability greater than $\frac{1}{2} + 1/|x|^c$. For statistical and perfect zero knowledge, we allow for unbounded distinguisher; equivalently, we require that the statistical difference between the distribution be negligible (less than $1/|x|^c$ for any $c$).

The above definition requires the existence of a single universal simulator, $S$, for all possible (efficient) verifiers. At first glance, the limitations on $S$ may seem to force $S$ to be as powerful as a prover. However, $S$ has important advantages over a prover $P$, allowing it to perform simulations in probabilistic polynomial time. First, it may set $\hat{V}$'s coin tosses as it wishes, and even run $\hat{V}$ on different sets of coin tosses. More importantly, $S$ may conceptually "back up" $\hat{V}$ to an earlier point in the conversation, and make different statements. This ability derives from $S$'s control of $\hat{V}$'s coin tosses; since $\hat{V}$ otherwise operates deterministically, $S$ can rerun it from the beginning, exploring different branches of the conversation tree.

Indeed, all known proofs of zero-knowledge construct black-box simulations. There is no way known to make use of a verifier's internal state, nor to customize simulators based on the description of $\hat{V}$ other than by using it as a black box.[1] Thus, given the current state of the art, an impossibility result for black-box zero-knowledge seems to preclude a positive result for the older definitions of zero-knowledge.

## 2.2 Black-box verifiers with private random functions

In this paper, as in [16], we note that it is enough to consider deterministic $\hat{V}$, i.e., even deterministic (yet, cheating) verifiers are hard to simulate. Also, following [16], we consider verifiers $\hat{V}$ that have access to a private random hash function $H$, that is wired into them and is not directly accessible to the simulator (note that $\hat{V}$ is deterministic in that it doesn't use an external source of randomness; its *construction* is randomized). That is, the simulator may gain only indirect access to $H$, by observing $\hat{V}$'s input/output behavior. For convenience, we assume that for any polynomially bounded $n$ and $m$, $H$ will take an $n$-bit input and return an $m$-bit output. In practice, $H$ will be defined for big enough $n$ and $m$, and its inputs (if short) will be padded to fit the length of $H$'s inputs. Pedantically, we can view $H$ as a family $\{H_{m,n}\}$ of hash functions; we suppress these subscripts for clarity.

As in [16], we will think of $H$ as being randomly chosen from a family of hash functions [7]. And as in [16] we do not use the standard pairwise independent family. Instead we use families of hash functions that achieve $p(n)$-independence, for some sufficiently large polynomial $p$. A member $H$ in this family can be described by a string of polynomial length, and it is this string that is wired into the verifier. The polynomial $p$ is set to exceed the running time of the simulator times the length of $V$'s answers. Thus, even if the simulator poses to $V$ a different query in each of its steps, and if for each query $V$ generates the hash of the query, using $H$, then the simulator will face a verifier that uses a completely random string for each of its (different) queries. Of course, if the simulator repeats a query, then the "deterministic" $V$ repeats the same response. Our use of the hash function will be as follows. The deterministic $\hat{V}$ will invoke the honest $V$ with a random tape determined by a hash on the history of the interaction so far (a history from interactions that do not involve $V$ and do not influence $V$'s actions normally). Thus, although $\hat{V}$ is deterministic, the hash of the history so far will give it the randomness that will foil the simulation.

---
[1] As one slight exception, [21] proves security against space-bounded verifiers by considering the internal state of the verifiers. However, these techniques do not seem applicable to more standard classes of verifiers.

We remark that the specification of H need not be a long polynomial string as assumed in the analysis. Instead, one may use a short seed and use a pseudo-random generator to choose a random H from the family of hash functions. This does not change the probability of failure for the prover we build by more than a negligible fraction.

## 2.3 Restricted interactive proofs

In this paper, we consider interactive proofs in which it is possible to tell whether $V$ will accept based on its conversation, without looking at its random coins. All the proofs we are aware of have this "conversation-based" property. The theorem is also correct without making this assumption. However, showing the theorem without the assumption requires a longer and more complicated proof. We feel that this complication is of small interest and we do not include it in the paper.

# 3 Creating an efficient prover

To prove our main theorem, we construct a particular malicious verifier (pedantically a family of closely related malicious verifiers), with a fixed scheduling strategy (a very similar strategy is used in [10]). We show that a simulator that successfully simulates the multi-conversation on input $x$ with high probability can be converted to a probabilistic polynomial prover $P_S$ for the original protocol. This prover will cause $V$ to accept $x$ with probability strictly greater than $1/2$. Thus, we can use this prover to probabilistically decide whether $x \in L$, implying that $L \in BPP$.

## 3.1 The attack

Let the original protocol consist of an initial challenge $q$, followed by a reply, $r$ and a second challenge, $s$ and a final reply $t$. The honest verifier $V$ generates $q$ as a function $q(x, R)$ of the input and its random coin flips, $R$. $V$ generates $s$ as a function $s(x, R, r)$ of the input, $R$ and $r$ ($q$ is implicit given $x$ and $R$). Finally, $V$ computes a predicate $accept(x, q, r, s, t)$ to determine whether to accept or reject. Note that this restricts the acceptance predicate to being "conversation-based," in which one can tell whether $V$ will accept based on its conversation, without looking at its random coins. As mentioned in Section 2.3 we prove our result with respect to such interactive proofs.

Let $k$ and $m$ be parameters that will be chosen later. (Both polynomial in the length of the input.) We consider the protocol obtained by performing $m$ proofs in parallel. Thus, we denote the initial challenge by
$$\vec{q} = (q^1, \ldots, q^m) = (q(x, R^1), \ldots, q(x, R^m)),$$
where $\vec{R} = (R^1, \ldots, R^m)$. We define $\vec{r}, \vec{s}$ and $\vec{t}$ analogously. Finally, we define $accept(x, \vec{q}, \vec{r}, \vec{s}, \vec{t})$ to be true iff $accept(x, q^i, r^i, s^i, t^i)$ for all $i$, $1 \leq i \leq m$. We call such a parallel set of proofs an $m$-block.

Note that parallel repetition is a special case of scheduling in an asynchronous environment. We remark that parallel repetitions reduce the error probability for interactive proofs but not necessarily for arguments (see [2]). ¿From now and on, we always use parallel repetitions, i.e., $m$-blocks proofs, instead of running a single message in each round.

Our attacking verifier $\hat{V}$ is defined by the value of its private random hash function, $H$ and parameters $k$ and $m$. $\hat{V}$ runs a total of $k$ $m$-block-proofs with the prover. We use subscripts to denote the version, e.g., $\vec{q}^i$ denotes the first question in the $i$th run of the protocol. (Remember that these are actually $m$ parallel first questions.) The verifier interleaves its challenges so that the sequence of messages appears as

$$\vec{q}_1, \vec{r}_1, \vec{q}_2, \vec{r}_2, \ldots, \vec{q}_k, \vec{r}_k, \vec{s}_k, \vec{t}_k, \vec{s}_{k-1}, \vec{t}_{k-1}, \ldots, \vec{s}_1, \vec{t}_1.$$

In the rest of the paper we fix this specific schedule, and assume the interaction is according to this schedule. Thus, our lower bound is valid even if the schedule is statically fixed and known to the simulator. The adversarial verifier $\hat{V}$ invokes the honest verifier $V$ for each of the proofs. However, the random tapes of the honest verifier are determined using the hash function, by

$$\vec{R}_i = H(x, \vec{q}_1, \vec{r}_1, \ldots, \vec{q}_{i-1}, \vec{r}_{i-1}). \tag{1}$$

Namely, it is the output of the hash function on the history of the interaction (of all concurrent proofs) so far. We assume that $H$ returns the correct number of random bits used by $V$. (In particular, $m$ random strings are required for each of the blocks.) The questions $\vec{q}_i$ and $\vec{s}_i$ are defined by: $\vec{q}_i = \vec{q}(x, \vec{R}_i)$ and $\vec{s}_i = \vec{s}(x, \vec{R}_i, \vec{r}_i)$. However, if for any $i$, $accept(x, \vec{q}_i, \vec{r}_i, \vec{s}_i, \vec{t}_i)$ does not hold, then $\hat{V}$ aborts immediately, without sending $\vec{s}_{i-1}, \ldots, \vec{s}_1$ to the prover.

We can thus view a conversation as consisting of two phases. The generation of

$$\vec{q}_1, \vec{r}_1, \vec{q}_2, \vec{r}_2, \ldots, \vec{q}_k$$

constitutes the *creation* phase, in which new blocks (values for $\vec{R}_i$) are created by $\hat{V}$, and the generation of

$$\vec{r}_k, \vec{s}_k, \vec{t}_k, \vec{s}_{k-1}, \vec{t}_{k-1}, \ldots, \vec{s}_1, \vec{t}_1$$

constitutes the *resolution* phase, in which these proofs run their course. We treat these phases quite differently when discussing the simulator.

Note that all the randomness used by $V$ as invoked by $\hat{V}$ comes from $H$. In particular, $\hat{V}$ is deterministic and doesn't use its random input, to some extent limiting the simulator $S$'s power over it.

## 3.2 The simulator

Our proof is by contradiction: assuming a simulator $S$ that properly handles our adversarial verifier $\hat{V}$ we construct an efficient prover $P_S$. The prover will use the simulator while feeding it with $\hat{V}$'s answers somewhat twisted. The ability of the simulator to simulate the interaction of the prover with $\hat{V}$ allows $P_S$ get the required information to convince the honest verifier on inputs in the language. This means that we have an efficient prover $P_S$ that convinces an efficient honest verifier $V$ on inputs in the language and fails to convince $V$ on inputs not in the language (by the soundness property of $V$). So we get an efficient procedure to determine if $x \in L$ and we are done.

In order to use the simulator by the prover $P_S$ we make some assumptions (without loss of generality) about the simulation, and give a convenient way of looking at the workings of the simulator. Our view of the simulator is especially tailored for the specific $\hat{V}$ described in the previous section. Recall that $\hat{V}$ invokes the honest $V$ with a random tape $\vec{R}$ that is determined by hashing of the conversation so far. Also, $\hat{V}$ proceeds with an interaction only if all other concurrent interactions that have ended, ended accepting.

First, we assume that whenever $S$ generates a transcript it runs it through $\hat{V}$. That is, before returning
$$(\vec{q_1}, \vec{r_1}, \vec{q_2}, \vec{r_2}, \ldots, \vec{s_2}, \vec{t_2}, \vec{s_1}, \vec{t_1}),$$
$S$ runs $\hat{V}$ to obtain $\vec{q_1}$, sends $\vec{r_1}$ to $\hat{V}$, receiving $\vec{q_2}$, and so on. Clearly, any simulator can be modified to perform in this manner without changing the quality of its simulation. Since our protocols are dialog based, we also assume without loss of generality that the simulator never sends $\hat{V}$ a value for $\vec{t_i}$ that would cause it to reject, since it could simply compute for itself that $\hat{V}$ would reject.

### 3.2.1 The proof tree

We now turn into representing the run of the simulator $S$ with a given verifier (and the predetermined schedule) as a tree. As it interacts with $\hat{V}$, $S$ implicitly generates many partial conversations, not all of which are successfully finished (only one need be). In generating these partial conversations, $S$ typically starts many proofs that may or may not be completed as the simulation proceeds. To discuss this interaction, we visualize these proofs using a leveled tree, which we call the *proof tree*. Each vertex $v$ of the graph corresponds to a new $m$-block that has been initiated between $S$ and $\hat{V}$. A vertex $v$ has parameters $\vec{R}, \vec{q}, \vec{r}$, corresponding to the beginning of a conversation. A vertex on level $i < k$ may have zero or more children on level $i+1$; the proof tree has at most $k$ levels. A vertex $v$ labeled $\vec{R_{i+1}}, \vec{q_{i+1}}, \vec{r_{i+1}}$ is a descendant of a vertex labeled $\vec{R_i}, \vec{q_i}, \vec{r_i}$ if the simulator runs $\hat{V}$ on a history in which $\hat{V}$ uses $\vec{R_i}, \vec{q_i}$ for the $i$-th proof, $S$ responds with $\vec{r_i}$, $\hat{V}$ then uses $\vec{R_{i+1}}, \vec{q_{i+1}}$ to continue the interaction, and $S$ responds with $\vec{r_{i+1}}$. We adopt the convention that Level 1 of the graph is the "top" level and Level $k$ is the bottom level.

Whenever $S$ runs a partial conversation
$$\vec{q_1}, \vec{r_1}, \ldots, \vec{q_i}, \vec{r_i}, \ldots$$
through $\hat{V}$, for $i \leq k$, it may be thought of as traversing/creating the proof tree as follows. $S$ will visit or create a sequence of vertices $v_1, \ldots, v_i$, where $v_j$ is on level $j$. First, $S$ visits or creates the top level vertex $v_1$ with parameters $(\vec{R_1}, \vec{q_1}, \vec{r_1})$. The values $\vec{R_1}, \vec{q_1}$ are determined by the verifier (according to the conversation so far, which is null) and the value $\vec{r_1}$ is then determined by the simulator. Then, for $1 < j \leq i$, after visiting/creating $v_{j-1}$, $S$ visits/creates $v_j$, the unique child of $v_{j-1}$ with parameters $(\vec{R_j}, \vec{q_j}, \vec{r_j})$. Again, $\vec{R_j}$ and $\vec{q_j}$ are sent by the verifier and the value $\vec{r_j}$ is a message that the simulator sends.

Note for example, that all siblings vertices will have the same values for $\vec{R}$ and $\vec{q}$, since they are all determined by the conversation so far. To avoid special cases, we adopt the convention that the top level vertices are siblings. Namely, we add a designated root vertex at level zero to the forest created so far and make it a tree by connecting the root to all level-1 vertices.

The second type of partial conversation simulated are those that contain also messages of type $\vec{s_i}$ and $\vec{t_i}$. Namely, after $k$ concurrent protocols have begun, these $k$ protocols are being continued

and finished. A simulated partial conversation of the form

$$\vec{q}_1, \vec{r}_1, \ldots, \vec{q}_k, \vec{r}_k, \vec{s}_k, \vec{t}_k, \ldots, \vec{s}_i, \vec{t}_i, \ldots$$

for $1 \leq i \leq k$ may be thought of as follows. First, the simulator traverses/creates a path from the top level to some bottom-level vertex $v$, with parameters $(\vec{R}_k, \vec{q}_k, \vec{r}_k)$, of the proof tree. At this point, it has simulated $\vec{q}_1, \vec{r}_1, \ldots, \vec{q}_k, \vec{r}_k$. When it runs $\hat{V}$ further, receiving $\vec{s}_k = \vec{s}(x, \vec{R}_k, \vec{r}_k)$, it is said to *activate* the vertex $v$. When it sends an acceptable $\vec{t}_k$ to $\hat{V}$, it is said to *resolve* $v$. Note that by our conventions $\vec{q}_k, \vec{r}_k, \vec{s}_k, \vec{t}_k$ is an accepting conversation. Thus, $v$ is resolved by the simulator $S$ when $S$ finds a message $\vec{t}_k$ that convinces $\hat{V}$ to accept and proceed to sending $\vec{s}_{k-1}$. When the simulator $S$ receives $\vec{s}_{k-1}$ it conceptually activates $v$'s parent, and so on. $S$ thus may be thought of as retracing its path back up the proof tree. The simulation may, without loss of generality be viewed as a series of such bounces. In general, $S$ may retrace partially, then continue down another path - but insisting that $S$ "start over" from the top level of the tree does not impair $S$'s efficiency (by more than a polynomial factor) or correctness (since $\hat{V}$ is deterministic).

We remark that two different nodes in the proof tree may have the same values in them. This may happen when two different histories result in the same $\vec{R}$ as determined by the hash function of $\hat{V}$. However, these nodes are not the same. Each has a history of interaction between the simulator and verifier as determined by the path from the root to the vertex. This event happens with negligible probability because of the randomness obtained from the hash function. However, when this event happens, it bears no effect on the proof.

## 3.3 Turning a simulator into a prover: Overview

We start with a high-level overview of how to convert a good simulator $S$ into an efficient prover $P_S$. The prover $P_S$ will run $S$. When $S$ asks to question $\hat{V}$'s behavior, $P_S$ will provide $\hat{V}$'s answers. Finally, $P_S$ will use the interaction of $S$ with $\hat{V}$ to convince the original honest verifier $V$ that $x \in L$.

We can view our $(S, \hat{V})$ interaction as generating and playing (not always to completion) many different proofs of the original protocol. On a very high level, our efficient prover $P_S$ "splices in" its interaction with $V$ into $\hat{V}$'s messages to the simulator $S$ and uses the answers created by the simulator as its own. This slicing operation cannot be noticed by the simulator $S$ since the slicing operation can be thought of as a random modification of the hash function: one specific value of going to be modified in a random manner (since the honest verifier chooses its random coins uniformly at random). The simulator will not be able to tell between a random hash function and a randomly modified hash function.

When the honest verifier $V$ chooses a random $R$ and send its first challenge, $q$, to $P_S$, $P_S$ conceptually alters the random hash function $H$ so that for some set of siblings in the proof tree, all having parameters

$$(\vec{R} = (R^1, \ldots, R^m), \vec{q} = q^1, \ldots, q^m, \cdot),$$

$R^j = R$ for some $j$, $1 \leq j \leq m$, and hence $q^j = q$. One of these siblings $v$ in particular, with parameter $(\vec{R}, \vec{q}, \vec{r})$ will be chosen. Under the right circumstances, $P_S$ will send $V$ the value of $r = r^j$; great care must be taken to actually send this $r$ to the honest verifier. This will happen later in the simulation, when the sliced vertex $v$ is activated. In order to generate $\vec{s}$; $P_S$ will send $r$ to $V$, receive $s$ and splice in $s^j = s$. Hopefully, $S$ will eventually resolve the vertex $v$ and respond with an acceptable $\vec{t}$; allowing $P_S$ to forward $t = t^j$, causing $V$ to accept.

Any splicing attempt may have three possible outcomes:

- $P_S$ can succeed in making $V$ accept,
- $P_S$ can fail (get stuck), losing its chance to make $V$ accept, or
- $P_S$ can abort, without causing $V$ to accept, but giving it the chance to try another slicing attempt.

Nearly all the time, $P_S$ will abort. With some small probability, $P_S$ will succeed and with some (hopefully much smaller) probability, $P_S$ will fail. We show that for most $R$, $P_S$ will succeed more often than it will fail, and can therefore try this splicing procedure repeatedly, ultimately succeeding with probability at least $2/3$.

There are a number of difficulties with the above approach. First, $S$ doesn't know $R$, so it can't really alter $H$ as described, and will have to check exactly how it can usually simulate the behavior of the spliced $\hat{V}$. A more technically problematic difficulty is that the simulator may generate many proofs that it never completes. Indeed, the ratio of completed proofs to uncompleted proofs may be quite small (though non-negligible). If $P_S$ sends $\vec{r}_v$ to $V$, and $S$ fails to generate an acceptable $\vec{t}_v$, $P_S$ will not be able to cause $V$ to accept.

To get around the problem of incomplete proofs, we have $P_S$ use a strategy that allows it to abort a splicing attempt before it has responded to $V$. We also have $P_S$ choose where to insert the real proof into the proof tree according to a particular distribution. We show that using these two techniques, $P_S$ may often abort but will only rarely fail.

**Remark:** If the original proof system has negligible error, then it suffices for $P_S$ to succeed with non-negligible probability. Given such an efficient $P_S$, one can run the proof many times to determine whether an input $x$ is in the language. Given an interactive proof with error bounded away from 1, one can run it in parallel to obtain a proof with negligible error. This gives a simpler proof of our theorem for this case, and can be used to extend the theorem to work for 5 message interactive proofs. Unfortunately, in the argument model, parallel amplification doesn't always work [2], so we can't use this trick to obtain a general theorem.

## 3.4 Splicing in the proof

Let us first state more explicitly how the splicing operation is done by $P_S$ when trying to prove to $V$ that $x \in L$. In the proof system, $V$ flips coins to generate $R$, and generates an initial challenge $q = q(x, R)$. The prover $P_S$ reads the value $q$ from its communication tape and invokes the simulator $S$ on input $x$ in the following manner. $P_S$ conceptually chooses a random hash function $H$, defining $\hat{V}$. Now $P_S$ can simply compute $H$ as needed on the fly. Thus, $P_S$ can exactly simulate the behavior of $\hat{V}$. $P_S$ then runs $(S, \hat{V})$, which starts generating the proof tree. Recall that if $S$ runs through the partial conversation $\vec{q}_1, \vec{r}_1, \ldots, \vec{r}_{i-1}$ and "asks" $\hat{V}$ for the next question $\vec{q}_i$, $\hat{V}$ then computes its parameters $\vec{R}$ and $\vec{q}$ by

$$\begin{aligned} \vec{R} &= H(x,_1, \vec{q}_1, \vec{r}_1, \ldots, \vec{q}_{i-1}, \vec{r}_{i-1}), \text{ and} \\ \vec{q} &= \vec{q}(x, \vec{R}). \end{aligned}$$

Also, recall that $\vec{R}$ and $\vec{q}$ are in fact a block of $m$ parallel repetitions of the actual proof as the proof that $P_S$ runs against the real verifier $V$.

Writing $\vec{R} = (R^1, \ldots, R^m)$, Define $splice(\vec{R}, j, R)$ to be the $m$-tuple equal to $\vec{R}$ at every coordinate accept the $j$th, and $R^j = R$. The splice operation takes place when $P_S$ randomly chooses $j$, $1 \leq j \leq m$ and conceptually modifies the function $H$ on one specific value:

$$H(x, \vec{q}_1, \vec{r}_1, \ldots, \vec{q}_{i-1}, \vec{r}_{i-1}) = splice(\vec{R}, j, R),$$

and hence $R^j = R$. Then, conceptually, $P_S$ just lets $S$ continue with the simulation (we show below the mechanics of how $P_S$ can do this).

The splice operation creates a set of siblings with the above values of $\vec{R}$ and $\vec{q}$. These are all the children vertices of the path in the tree whose value in the hash function has been modified. One of these siblings, $v$, may be chosen to generate the conversation with the original verifier $V$. This special sibling $v$ is chosen at random as will be described later.

If the special sibling $v$ is later activated, i.e., one of its children has been solved and the simulator expects to get $\hat{V}$'s message $\vec{s}$ in the proof represented by the vertex $v$, then the prover $P_S$ must generate $\hat{V}$'s response $\vec{s} = s^1, \ldots, s^m$. At this stage, $P_S$ cannot determine $S^j$ since $R^j$ is only known to $V$. $P_S$ is able to determine all $s^\ell$ for $1 \leq \ell \leq m$ and $\ell \neq j$. Thus, $P_S$ needs to get $s^j$ from its interaction with the honest verifier $V$. Therefore, at this specific time, $P_S$ sends $r = r^j$ to $V$, receives $s$ and sets $s^j = s$ so that it can continue the run of the simulator.

Here is where $P_S$ suffers from not knowing $R$, which would allow $P_S$ to compute $s^j$ by itself. It uses $V$ to perform this computation for it, but note that this trick may be performed only once (and at great risk). Once $P_S$ gave the honest verifier a message $r$ and got its response $s$, it is not possible for $P_S$ to change its mind and send $V$ a different message $r'$. The honest verifier does not agree to such "rewinding" of a proof. $P_S$ cannot waste this try it on a sibling of $v$, since it needs this try to activate the vertex $v$. So, if a sibling of $v$ is activated before $v$ is activated, $P_S$ aborts. Note that in this case, $P_S$ has a chance to try again, since no $r$ has been sent to the honest verifier $V$, and $V$ is still waiting for such an $r$ to appear. Trying again means randomly choosing a new $H$ and a splice index, and starting a new simulation with $S$ from scratch. The case that $P_S$ fails completely is when a sibling of $v$ is activated after $v$ is activated. In this case, $P_S$ has sent $V$ the message $r$, but is required to respond to the simulator on another sibling, namely, a different $r$. If that happens before the simulator resolves $v$, then $P_S$ cannot proceed. We assume in a worst case manner that if a sibling of $v$ is activated, then $P_S$ fails. We announce this failure even if the sibling is activated before $v$, or even if the sibling is activated after $v$ has already been resolved.

If $S$ resolves $v$, generating a value of $\vec{t} = t^1, \ldots, t^m$ that will cause the spliced $\hat{V}$ to accept, then $P_S$ sends $t$ to $V$. At this point, $P_S$ succeeds in convincing the honest verifier $V$ (Here we use the fact that $S$ only sends an acceptable $\vec{t}$).

Except for choosing where to splice in the proof, we have specified $P_S$. It can be verified that, at least up to the point where $P_S$ aborts, $P_S$ perfectly simulates the behavior of $\hat{V}$ with the spliced $H$ for the simulator. Namely, the view of the simulator $S$ in case it gets $\hat{V}$ (with a randomly chosen $H$) as a black box is identical to the view of the simulator in case it is invoked by $P_S$ with a spliced $\hat{V}$.

## 3.5 Choosing the spliced vertex $v$

We define the *height* of a vertex at level $i$ to be $h = k - i + 1$. Namely, a vertex that represents the first proof in the schedule has height $k$ and a vertex representing the $k$-th proof in the schedule has height 1. Our method for choosing which special vertex $v$ to splice in the conversation with $V$ obeys the following design criteria:

- The probability that a generated vertex $v$ is the special one depends only on its height, and is completely independent of any other aspect of the entire run of $S$.

- If $v$ has height $h$, it's probability will be proportional to $f(h)$, for some carefully chosen $f$.

First, let us bound the running time of the (average) polynomial time simulator. Let $n = |x|$ and let $\alpha$ be a constant such that $S$ runs in an expected number of steps at most $(nkm)^\alpha$. Since $P$ always causes $\hat{V}$ to accept, $S$ must cause $\hat{V}$ to accept (that is, generate a path going all the way down and then all the way up the proof tree) with probability close to 1. By Markoff's inequality, if we only allow $S$ to run for $N = 100(nkm)^\alpha$ steps, it will still succeed with probability greater than $0.9$. Indeed, its simulation will no longer be close to the actual one, but, our analysis will only consider whether $S$ causes $\hat{V}$ to accept. For the rest of the analysis, $S$ will only run for $N$ steps.

With $S$ as above, every level has at most $N$ vertices. We use the following addressing scheme for the vertices of the proof tree. As discussed in Section 3.2.1, we imagine a dummy root vertex at Level 0 that has all the Level 1 vertices as children. For each vertex $v$ generated on level $i$ we keep track of how many vertices have been generated on the same level before $v$. To each vertex $v$ we assign an address $(i, a)$, denoting that $v$ is on level $i$ and is the $a$th vertex generated on level $i$. Note that the content of an address is a random variable determined by the run of the simulator with $\hat{V}$. In particular, it is possible that at some level $i$ there will be less than $N$ vertices and thus an address will not correspond to any actual vertex generated in the proof tree. However, each vertex that is actually generated has an address $(i, a)$ with $1 \leq i \leq h$ and $1 \leq a \leq N$. To choose the special vertex $v$ for the splicing, we choose the address of $v$ at random by first picking a level $i$ with probability $cf(h)$, where $h = k - i + 1$ and $c$ is the normalizing constant defined by $\sum_{h=1}^{k} cf(h) = 1$, and choosing $a$ uniformly subject to $1 \leq a \leq N$. Thus, any address of height $h$ is selected with probability $cf(h)/N$.

The precise function $f$ is a polynomial $f(h) = h^\beta$ chosen to make the analysis work out; we defer the determination of $\beta$ to that section. Note that this choice puts a higher probability weight on higher vertices.

## 4  Preliminary analysis of the splicing operation

We now bound below the probability that $P_S$ succeeds and the ratio of the probability that $P_S$ succeeds to the probability that $P_S$ fails when $R$ is chosen uniformly. As discussed later, this is *not* sufficient to prove our theorem, but is a very good start. We begin with formalizing the fact that the simulator $S$ cannot tell if (and where) the proof tree is being spliced. Then, we relate properties of the proof tree with the probability that the simulator succeeds or fails, and finally, we bound these probabilities by bounding probabilities of events that are related to the generation of the proof tree by the simulator. We first consider the following experiment.

EXPERIMENT 1: Execute the following steps:

1. Choose a random address $(i, a)$ as above.

2. Choose $R$ uniformly (over strings of the appropriate length) and run $V$ with $R$.

3. Generate and traverse the proof tree by running $S$ against $\hat{V}$ with a randomly chosen $H$, choosing $(i, a)$ as the address of the special vertex $v$, and splicing accordingly.

4. Output the traversed tree, the order in which the vertices in the tree were generated/traversed, and $(i, a)$.

Since we output the proof tree and the order in which vertices are generated/visited, we can determine which vertex is $(i, a)$, and for each vertex we can determine whether it has been activated or resolved. We also assume, for the sake of the analysis, that the simulation continues even if $P_S$ fails. That is, once we splice the game in, we allow $P_S$ to query the honest verifier $V$ multiple times. Of course, we have to take into account the fact that $P_S$ really failed in this case.

**Definition 2** *Given a run of the simulator with a verifier $\hat{V}$, we define the following properties of an address $(i, a)$ in the run of Experiment 1. These properties depend on the random coins of the simulator and $\hat{V}$'s messages. The messages of $\hat{V}$, in turn, depend on the random $H$ that it uses, the spliced address $(i, a)$ and the random coins of the real verifier $R$.*

1. *We say that an address $(i, a)$ in the output of experiment 1 is* good *iff the vertex at that address is resolved and no sibling vertex is ever activated.*

2. *We say that an address $(i, a)$ is* bad *iff the vertex at that address is activated but not resolved or if any sibling vertex is ever activated.*

3. *We say that an address $(i, a)$ is* interesting *if it is either good or bad.*

4. *We say that a vertex $v$ is good/bad/interesting if its address in the given run is good/bad/interesting.*

The following lemma follows from the construction of $P_S$ and the definitions.

**Lemma 2** *The probability that $P_S$ succeeds in convincing $V$ is at least the probability that the chosen address $(i, a)$ is good with respect to the run of $S$ against the spliced $\hat{V}$. The probability is taken over the random tape of $S$, the choice of $(i, a)$ for the slicing, the choice of the function $H$ determining $\hat{V}$'s messages, and the random tape $R$ of $V$. The probability that $P_S$ fails is bounded above by the probability that $(i, a)$ is bad with respect to the run of $S$ against the spliced $\hat{V}$.*

We now formalize the fact that the simulator cannot tell whether the interaction with $\hat{V}$ has been sliced or not. Recall that when $R$ is chosen uniformly, all the splicing operation does is conceptually replace a uniformly chosen value (of the hash function) with another uniformly chosen value. This is exactly the case for us, since the simulator cannot ask enough questions to tell that $H$ is not completely random. It asks at most $N$ questions and $H$ is $N$-wise random. It follows that the splicing of $(i, a)$ using the original verifier $V$ yields the same distribution on the run of the simulator (and thus, on how the proof tree is generated and traversed) and this distribution is exactly the same as if we never spliced in a game. We can therefore reorder the steps of the experiment as follows.

EXPERIMENT 2: Execute the following steps:

1. Generate and traverse the proof tree by running $S$ against $\hat{V}$ with a randomly chosen $H$ (with no splicing).

2. Choose a random address $(i, a)$ as above.

3. Output the traversed tree, the order in which the vertices in the tree were generated/traversed, and $(i, a)$.

Note that the notion of being activated and resolved is simply a property of the tree generation/traversal, and is therefore still well defined. We will now associate success and failure of the simulator with the type of vertices in the proof tree. Lemma 3 follows from the above discussion and a straightforward application of Bayes theorem.

**Lemma 3** *The probability that $P_S$ succeeds is bounded below by the probability that the output $(i, a)$ is good with respect to the tree output by Experiment 2. Here the probability is taken over the random tape of the simulator $S$, the choice of $(i, a)$ for the output, and the choice of the function $H$ determining $\hat{V}$'s messages. The probability that $P_S$ fails is bounded above by the probability that the output of Experiment 2 satisfies that the output $(i, a)$ is bad with respect to the proof tree in the output.*

The event that the output $(i, a)$ is good with respect to the output tree is equal to the sum over all good addresses $(i', a')$ in the tree of the probability that $(i, a) = (i', a')$ (and similarly for the probability that $(i, a)$ is bad). We can thus recast Lemma 3 as the following calculation. Consider the random variables SUCCEED, FAIL and INTERESTING, generated as follows. An $H$ is chosen at random for $\hat{V}$, and then $S$ (on a uniformly chosen random tape) generates and traverses the proof tree, and finally, it outputs the proof tree and we compute

$$\text{SUCCEED} = \sum_{(i,a) \text{ good}} cf(k-i+1)/N,$$

$$\text{FAIL} = \sum_{(i,a) \text{ bad}} cf(k-i+1)/N \text{ and}$$

$$\text{INTERESTING} = \sum_{(i,a) \text{interesting}} cf(k-i+1)/N.$$

Note that INTERESTING = SUCCEED + FAIL. We will bastardize terminology slightly, and speak of these variables after a given generation/traversal of a proof tree.

**Lemma 4** *The probability that $P_S$ will succeed is at least $E(\text{SUCCEED})$ and the probability that $P_S$ will fail is at most $E(\text{FAIL})$. The expectation is taken over the choice of the random tape for the simulator and the hash function $H$ for $\hat{V}$.*

To bound these expectations, we need to consider the structure of good and bad addresses.

### 4.1 The structure of bad and good addresses

Recall that a property of the verifier $\hat{V}$ is that it does not carry on with any of the proofs once one of them fails. Thus, $\hat{V}$ will never provide its second question (i.e., the third message of the proof) for a level-$i$ proof for $i < k$ unless one of its children (in the proof tree) has been completed and $\hat{V}$ accepts. As a consequence, any interesting (good or bad) vertex has to have a child vertex that is resolved (and also interesting). This is the main restriction that makes the life of the simulator difficult. It cannot get the verifier's third message of protocol $j + 1$ before it has resolved protocol

$j$, i.e., it has produced an answer that is convincing the verifier for protocol $j$. We phrase this restriction as a combinatorial property of the interesting (good and bad) addresses. We start by defining a snake. Loosely speaking, this is a path in the tree that goes down to a leaf.

**Definition 3** *A* snake *$\sigma$ is a series of vertices*

$$v_i, v_{i+1}, \ldots, v_k$$

*such that $v_j$ is on level $j$ and $v_{j+1}$ is a child of $v_j$ for $i \leq j < k$. We call $v_i$ the* head *of $\sigma$ and $v_{i+1}, \ldots, v_k$ the* body *of $\sigma$. We define the height of the snake $h(\sigma)$ to be $k - i + 1$ (this is the length of the snake and the height of its head).*

**Lemma 5** *For any generation/traversal of a proof tree by the simulator together with any $\hat{V}$, the set of interesting vertices can be canonically decomposed into disjoint snakes such that*

- *If a vertex is a bad vertex with no bad siblings then it must be a head of one of the snakes.*

- *Given a set of bad siblings, at most one of them is in the body of a snake. The rest of the bad siblings must be heads of snakes in the decomposition.*

**Proof:** Given a proof tree we build the set of snakes in the following manner. From each interesting vertex $v$ without an interesting parent, we start a snake, with $v$ as the head. If $v$ is on level $k$ we're done, else $v$ must have at least one resolved children, thus, it has at least one interesting child. In some canonical fashion, choose one of the interesting children to recursively continue the snake, and start new snakes at the other interesting siblings.

The properties required by Lemma 5 are easily verified: If a vertex is bad, then it is either activated but has activated siblings or it is activated but not resolved. Suppose the second case holds and the first does not. Then it is a bad vertex with no bad siblings. In this case, since the vertex has not been resolved, the verifier $\hat{V}$ has not activated its parent and this is an interesting vertex with no interesting parent. Thus, this vertex is a head of a snake as required. Suppose the first case holds, i.e., several siblings have been activated. If their parent is not interesting then all of them are heads of snakes (and no one is in a body of a snake). Otherwise, the parent in interesting and one of them at most is chosen to continue the parent's snake, whereas the rest are snake heads. Thus, again, at most one of these siblings is in the body of a snake as required. □

We would like to use this structure of good and bad vertices to show that the probability of choosing a bad vertex in the output is smaller than the probability of choosing a good vertex. We note that most of the bad vertices are heads of snakes. Actually for each bad vertex that is not a head of a snake, there exists at least one sibling that is a head of a snake. On the other hand, the interesting vertices are all vertices in all snakes. We will show that the weight of bad vertices is much smaller than the weight of interesting vertices. Thus, the weight of good vertices (interesting but not bad) is high enough.

Given a canonical snake decomposition of the interesting vertices of a graph, we can bound FAIL and INTERESTING as follows. Let $F(i) = \sum_{j=1}^{i} f(i)$.

**Lemma 6** *For any proof tree output by experiment 1 or 2,*

$$\text{FAIL} \leq 2 \sum_{\sigma} cf(h(\sigma))/N \text{ and}$$
$$\text{INTERESTING} \geq \sum_{\sigma} cF(h(\sigma))/N.$$

**Proof:** To establish the bound for FAIL, note that $cf(h(\sigma))/N$ is simply $cf(k-i+1)/N$, where $i$ is the level of $\sigma$'s head. By Lemma 5, A lone bad vertex (without a bad sibling) must be at a head of a snake and is therefore counted in the summation. By Lemma 5, given a set of bad siblings, all of them but one are heads of snakes and are thus counted. The worst case of undercounting the bad vertices is when there are exactly two bad siblings of which one is the head of the snake but the other isn't. However, this undercounting is compensated by the 2 in front of the summation. This establishes the bound for BAD. To establish the bound for INTERESTING, note that $cF(h(\sigma))/N$ simply sums the contribution of each vertex in the snake. Since the snakes are disjoint, the weight of all interesting vertices is the sum of weights of all vertices in all snakes. □

The above lemmas connect the snakes decomposition to the weight of good and bad vertices. Each snake decomposition satisfies these requirements. We would like now to go on and show that there is higher weight on good vertices than on bad vertices when the simulator succeeds in outputting a conversation. We will first note that long snakes are good for us, and then note that all the short snakes are overshadowed by one full snake of a successful simulation.

## 4.2 Setting the parameters

We first set $k = |x|$, i.e., the number of block proofs run by the schedule is the length of the input. We choose the number of parallel proofs in each proof block to be $m = k^3$. Recall that $N$ is a bound on the running time of the simulator, which is polynomial in the length of the input. We choose $\beta$ to be around the degree of a that polynomial. Specifically: $\beta \stackrel{\text{def}}{=} 1 + \lceil \log_k(N) \rceil$. Finally, we set $f(h) = h^\beta$, and thus $c$, the normalizing factor, is $c = 1/\left(\sum_{h=1}^{k} h^\beta\right)$. It is easy to verify that for any $h$ and any positive integer $\beta$, $F(h) \geq h^{\beta+1}/(\beta+1)$, and hence $F(h)/f(h) \geq h/(\beta+1)$.

We would like now to distinguish long snakes (which will be good for us) from short snakes (which will be dominated by the long ones).

**Definition 4** *We say that a snake $\sigma$ is* short *if $h(\sigma) < 10(\beta+1)$ and* long *otherwise.*

Using Lemma 6 and the fact that $F(h(\sigma))/f(h(\sigma)) > 10$ for a long snake $\sigma$, we have

$$
\begin{aligned}
\text{INTERESTING} &\geq \sum_{\text{long } \sigma} cF(h(\sigma))/N \\
&\geq 5\left(\sum_{\text{long } \sigma} 2cf(h(\sigma))/N\right) \\
&\geq 5\left(\text{FAIL} - \sum_{\text{short } \sigma} 2cf(h(\sigma))/N\right) \\
&\geq 5\left(\text{FAIL} - \sum_{\text{short } \sigma} 2c(10(\beta+1))^\beta/N\right) \\
&\quad [\text{Since } \sigma \text{ is short}] \\
&\geq 5\left(\text{FAIL} - 2c(10(\beta+1))^\beta\right)
\end{aligned}
$$

The last inequality follows because there are at most $N$ snakes in the decomposition. Summing up, we get:
$$\text{FAIL} \leq \text{INTERESTING}/5 + 2c(10(\beta+1))^\beta. \tag{2}$$

Now, if the constant term were 0, this would imply that we succeed much more than we fail, since INTERESTING = SUCCEED + FAIL. We would like to show that the constant term is small comparing to INTERESTING and we will do that by using the fact that there is a long snake: the one that is output by the simulation. However, the long snake appears only if the simulation ends well. So we need to compute the expected value of the constant term over the random coins of the simulator. Recall that it succeeds with probability at least 0.9. By the linearity of expectation we get that for any value of $H$ determining the behavior of $\hat{V}$,

$$E(\text{FAIL}) \leq E(\text{INTERESTING})/5 + 2c(10(\beta+1))^\beta. \tag{3}$$

Where the expectation is taken over the random coins of the simulator running against the fixed verifier $\hat{V}$. We now show that the constant term is small comparing to $E(\text{INTERESTING})$.

**Lemma 7** *Fix a hash function $H$ and a corresponding verifier $\hat{V}$, then*

$$2c(10(\beta+1))^\beta \leq E(\text{INTERESTING})/10,$$

*where the expectation is taken over the random coin tosses of the simulator.*

**Proof:** Recall that the simulator must work on $x \in L$ against any possible verifier $\hat{V}$. Note also that in the original interaction $\hat{V}$ is always convinced by the prover. Thus, with probability at least $0.9$, $S$ succeeds in finishing all $k$ games and then there is a snake of height $k$ (from the bottom to the top). Also, recall that $\beta$ was set so that $N < k^{\beta-1}$; therefore:

$$\begin{aligned} E(\text{INTERESTING}) &\geq 0.9cF(k)/N \\ &\geq 0.9c\frac{k^\beta}{k^{\beta-1}} = 0.9ck \end{aligned}$$

For sufficiently long inputs (and thus sufficiently large $k$'s), $0.9k$ is greater than the constant $20(10(\beta+1))^\beta$ and we are done. □

Finally, Lemma 8 follows from Lemma 7, Equation 3 and the fact that $c$ is a polynomial fraction in $k$.

**Lemma 8** *$P_S$ succeeds with probability at least $1/k^\gamma$ for some constant $\gamma$. Furthermore, $P_S$ succeeds at least 4 times as often as $P_S$ fails.*

## 5 Showing success for most $R$

Naively, one might suppose that Lemma 8 would imply that we are done. Given an input $x \in L$, $P_S$ keeps on trying the splicing strategy (with the parameters determined above) until it succeeds or fails. It will conclude in expected polynomial time and it will succeed with probability at least $4/5$, implying that $L \in BPP$. However, this analysis would only hold if $V$ chose $R$ independently for each of $P_S$ attempts; in reality, $V$ chooses $R$ once. The problem is that the previous section bounds the expected success rates over a random simulator coin, and thus over a random $R$. Also, note that it is possible that only a small fraction of the vertices traversed by the simulator are interesting, and the simulator has some control over which of them are. So if we only have Lemma 8 it is possible that for a very small fraction of $R$, $P_S$ succeeds much more often than it fails, yet for the rest of the $R$, $P_S$ fails a bit more often than it succeeds. Thus, Lemma 8 is not enough.

We now use the parallel repetitions to finish the proof. Recall that each vertex in the proof tree corresponds to a block of $m = k^3$ parallel copies of the original proof. Out of these $m = k^3$ original proofs at most one is altered by the splicing operation. Loosely speaking, we exploit this fact to show that the conditioning over one random entry in such a big block of proofs is, in a way, unnoticeable and the chances of hitting a good vertex, even when $R$ is "almost fixed" is still higher than the chances of hitting a bad vertex. More formally, we will show that for nearly all $R$, the success and failure probabilities are close to the expected values over the random tape of the simulator as stated in Lemma 8.

**Lemma 9** *For all but measure .01 of the random strings $\mathcal{R}$, $P_S$ succeeds with probability $\frac{1}{2k^\gamma}$ for some constant $\gamma$. Furthermore, for all but measure .01 of the random strings $\mathcal{R}$, $P_S$ succeeds at least 3 times as often as $P_S$ fails.*

**Remark:** By setting parameters correctly, the 3 can be replaced by anything less than the corresponding value in Lemma 8. In turn, this value (set arbitrarily at 4) can be set to any constant (or indeed, can be polynomially large, with care). Similarly, the .01 may be made arbitrarily (though non-negligibly) small.

We will prove Lemma 9 using Lemma 10 and 11 below. But let us first explain why Lemma 9 is sufficient for proving our result.

**Proof of Theorem 1:** As described above, we build a simulation based prover $P_S$ that can convince the honest verifier to accept with probability $2/3$ on $x \in L$. For $x \notin L$ $P_S$ convinces $V$ with probability at most $1/2$ by the soundness property of $V$. Thus we get a probabilistic polynomial time algorithm to determine $L$: Run $P_S$ against $V$ and accept iff $V$ accepts. Since both $P_S$ and $V$ are efficient, this can be done in polynomial time.

It remains to show that $P_S$ succeeds with probability $2/3$ when $x \in L$. We use Lemma 9. With probability .02 a bad random tape is chosen for $V$ and then $P_S$ looses. But if $V$ chooses a good random string, the $P_S$ will succeed in the following procedure.

Choose uniformly at random a random tape for the simulator and a hash function $H$. Run the simulator against the verifier $\hat{V}$ with the chosen $H$, while splicing the real interaction with $V$ into the interaction of the simulator with $\hat{V}$ as described above. By Lemma 9, $P_S$ succeeds with probability $p \geq \frac{1}{k^\gamma}$, fails with probability $q \leq p/3$ and gets another chance with probability $1 - p - q$. Thus, the simulator can repeat the experiment for $k^{3\gamma}$ times unless it fails or succeeds. Success happens with probability at least $2/3$ as required. □

It remains to prove lemma 9. We start by showing that if one random entry in an $m$-length random $R$'s vector is modified to contain a random $R$ chosen out of a small (but not too small) subset of the $R$'s, then this doesn't affect the distribution space too much. Let $\mathcal{R}$ denote the set of $V$'s possible coin tosses and $\mathcal{R}'$ denote an arbitrary subset of $\mathcal{R}$ of measure at least .01 (that is, a uniformly chosen $R$ is in $\mathcal{R}'$ with probability at least .01). Let $D$ denote the uniform probability measure on $\mathcal{R}^m$ and let $D'$ denote probability measure obtained by choosing $(R^1, \ldots, R^m)$ uniformly, choosing $j$, $1 \leq j \leq m$ uniformly, and then replacing $R^j$ by a uniformly chosen element of $\mathcal{R}'$. Lemma 10 says that these measures are multiplicatively close over most elements of their domain.

**Lemma 10** *For all $R$'s except for a negligibly small measure (over both $D$ and $D'$) it holds that:*
$(1 - \frac{1}{k}) Pr_D(\vec{R}) \leq Pr_{D'}(\vec{R}) \leq (1 + \frac{1}{k}) Pr_D(\vec{R}).$

**Proof:** Suppose that $\vec{R}$ has $\ell$ elements in $\mathcal{R}'$ and let $\rho$ denote the measure of $\mathcal{R}'$ under the uniform measure. By an elementary probability argument, we can explicitly derive the relevant probabilities, obtaining

$$Pr_D(\vec{R}) = \frac{\binom{m}{\ell}\rho^\ell(1-\rho)^{m-\ell}}{|\mathcal{R}'|^\ell|\mathcal{R}|^{m-\ell}} \text{ and}$$

$$Pr_{D'}(\vec{R}) = \frac{\ell\binom{m}{\ell}\rho^{\ell-1}(1-\rho)^{m-\ell}}{m|\mathcal{R}'|^\ell|\mathcal{R}|^{m-\ell}}$$

Thus, we have

$$\frac{Pr_{D'}(\vec{R})}{Pr_D(\vec{R})} = \frac{\ell}{\rho m}.$$

Thus, the inequality of Lemma 10 holds as long as

$$(1-\frac{1}{k})\rho m < \ell < (1+\frac{1}{k})\rho m. \tag{4}$$

It remains to bound the probability that $\ell$ deviates significantly from $\rho m$. But note that by the distribution $D$ this is just the tail of the Binomial distribution and can be bounded using the Chernoff bound. Recalling that $m = k^3$ and that $\rho > 0.1$, we get that Equation 4 holds with probability close to 1 (up to an exponentially small (in $k$) fraction) under the distribution $D$. It remains to note that $D'$ deviates from $D$ by at most one entry. Thus, the difference in the value of $\ell$ is at most 1, and thus Equation 4 holds under $D'$ with essentially the same probability. $\square$

**Definition 5** *We denote by $\mathcal{B}$ the set of negligible measure of random tapes $\vec{R}$ for which Equation 4 does not hold (and thus, the guarantee in Lemma 10 does not necessarily hold).*

We go on and show a similar result which is more specific to our setting. We define two probability spaces over the hash functions. The first distribution is the uniform distribution over all $H$'s in the family and we denote this probability space by $\mathcal{H}$. The second set of distributions is indexed by an address $(i, a)$, a random tape $s$ for the simulation, and a set $\mathcal{R}'$. We denote this distribution by and denoted $\mathcal{H}(i, a, s, \mathcal{R}')$ depends on the behavior of the simulator and on any arbitrary predetermined set of random coins $\mathcal{R}'$ of measure at least $.01$. It is defined by the following sampling procedure.

- Choose $H$ by $\mathcal{H}$.

- Choose an index $1 \leq j \leq m$ uniformly at random.

- Choose $R \in \mathcal{R}'$ uniformly at random.

- Run the simulator $S$ on the random tape $s$ with the black box being the verifier $\hat{V}$ with $H$. If a vertex $(i, a)$ appears, modify one entry in $H$ so that vertex $(i, a)$ contains $R$ in its $j$-th location. Output the resulting $H$. Otherwise (if a vertex at address $(i, a)$ does not appear), output $H$ with no modifications.

All above random choices are independent of each other.

Note that if we fix a random tape $s$ to the simulator $S$ and a hash function $H$, then the proof tree is fixed. Also, for each given $s$, if $H$ is chosen uniformly at random, then the probability that the proof tree includes a random tape $\vec{R} \in \mathcal{B}$ is negligible. This is because the tree contains a polynomial number of $\vec{R}$'s, which, for a random $H$, are selected uniformly at random. Denote the set of such pairs $(H, s)$ that have an $R \in \mathcal{B}$ in their proof tree by $\mathcal{C}$. The set $\mathcal{C}$ has negligible probability to be picked when $s$ and $H$ are uniformly chosen. But note that this set also has negligible probability when $H$ is chosen under $\mathcal{H}(i, a, s, \mathcal{R}')$ for any possible $i, a, s$. In all vertices except for the modified $(i, a)$, $H \in \mathcal{H}(i, a, s, \mathcal{R}')$ behaves uniformly. And at the vertex $(i, a)$ only one entry out of the $m$ gets modified. So the probability that the resulting $\vec{R}$ is in $\mathcal{B}$ is still negligible.

**Lemma 11** *For any $\mathcal{R}'$ of measure at least .01 and for all pairs $(H, s)$ not in $\mathcal{C}$ it holds that for all addresses $(i, a)$, $1 \leq i \leq k$ and $1 \leq a \leq N$,*

$$(1 - \frac{1}{k})Pr_{\mathcal{H}}(H) \leq Pr_{\mathcal{H}(i,a,s,\mathcal{R}')}(H) \leq (1 + \frac{1}{k})Pr_{\mathcal{H}}(H)$$

**Proof:** We concentrate on the right Inequality in Lemma 11. The left Inequality follows in the same manner. If $(i, a)$ does not appear in the interaction of $S_s$ with $\hat{V}_H$, then $Pr_{\mathcal{H}(i,a,s,\mathcal{R}')}(H) = Pr_{\mathcal{H}}(H)$ and we are done. (Note that the modification of $H$ at address $(i, a)$ in the sampling procedure of $\mathcal{H}(i, a, s, \mathcal{R}')$ does not make address $(i, a)$ appear or disappear from the proof tree.) Suppose $(i, a)$ does appear in the proof tree traversed by $\hat{V}_H$ with $S_s$, and consider the sampling procedure of $\mathcal{H}(i, a, s, \mathcal{R}')$. First, a function $H$ is chosen by $\mathcal{H}$, and then if the entry $(i, a)$ appears in the run of $\hat{V}_H$ with $S_s$ then a random entry $1 \leq j \leq m$ is chosen in that vertex and a random $R'$ is chosen in $\mathcal{R}'$ and then the hash function is modified so that its value at vertex $(i, a)$ contains $R'$ in the $j$th entry. Denote by $\vec{R}_{i,a}$ the vector $\vec{R}$ that appears in vertex $(i, a)$ before the modification with the newly chosen $R'$. We may write the probability of choosing an $H$ through this procedure by:

$$Pr_{\mathcal{H}(i,a,s,\mathcal{R}')}(H) = \sum_{\vec{R}} Pr_{\mathcal{H}}(\vec{R}_{i,a} = \vec{R}) \cdot Pr_{\mathcal{H}(i,a,s,\mathcal{R}')}(H \mid \vec{R}_{i,a} = \vec{R}) \quad (5)$$

Let us now bound $Pr_{\mathcal{H}(i,a,s,\mathcal{R}')}(H \mid \vec{R}_{i,a} = \vec{R})$. First, if vertex $(i, a)$ in the interaction of $\hat{V}_H$ with $S_s$ is different from $\vec{R}$ in more than one entry, then this probability is zero. What is the probability that a specific function $H$ is sampled by $\mathcal{H}(i, a, s, \mathcal{R}')$ given the $\vec{R}_{i,a} = \vec{R}$? It is the probability that all the entries other then $(i, a)$ were chosen as $H$ and that $\vec{R}$ was modified to fit our $H$. Let us compute this probability by summing over all choices of $H$'s (by $\mathcal{H}$) and $j$'s that may have led to picking this hash function. Note that once $H$ and $s$ are fixed up to one change in a specific vertex, the string on which $H$ is modified is also determined. Let this entry be $\alpha$ and $\vec{R}$ is $H'(\alpha)$ for $H'$ that was picked by $\mathcal{H}$ before the modification. Let $\mathcal{D}(H, \alpha, j)$ be the set of $H$'s that have the same values as our $H$ on all entries expect at the $j$th index on input $\alpha$. The cardinality of $\mathcal{D}(\mathcal{H}, \alpha, |)$ is exactly $|\mathcal{R}|$, the number of possible values that $R$ may take.

$$Pr_{\mathcal{H}(i,a,s,\mathcal{R}')}(H \mid \vec{R}_{i,a} = \vec{R}) = \sum_{j:\vec{R}_j \in \mathcal{R}'} \frac{1}{m} \cdot \frac{1}{|\mathcal{R}'|} \cdot \sum_{H' \in \mathcal{D}(H,\alpha,j)} Pr_{\mathcal{H}}(H')$$

Since the probability of all possible $H$'s are equal in $\mathcal{H}$, we get:

$$Pr_{\mathcal{H}(i,a,s,\mathcal{R}')}(H \mid \vec{R}_{i,a} = \vec{R}) = \frac{\ell}{m} \cdot \frac{|\mathcal{R}|}{|\mathcal{R}'|} \cdot Pr_{\mathcal{H}}(H)$$

where $\ell$ is the number of indices $1 \leq j \leq m$ satisfying that $\vec{R}'_j \in \mathcal{R}'$. Now, since the pair $(H, s)$ is not in $\mathcal{C}$, then $R'$ is not in $\mathcal{B}$, so $\ell \leq (1 + \frac{1}{k})m\frac{|\mathcal{R}'|}{|\mathcal{R}|}$, and

$$Pr_{\mathcal{H}(i,a,s,\mathcal{R}')}(H \mid \vec{R}_{i,a} = \vec{R}) \leq (1 + \frac{1}{k}) \cdot Pr_{\mathcal{H}}(H) \tag{6}$$

Using Equation 6 in Equation 5, we get:

$$Pr_{\mathcal{H}}(H) \leq (1 + \frac{1}{k}) \cdot \sum_{\vec{R}} Pr_{\mathcal{H}(i,a,s,\mathcal{R}')}(\vec{R}_{i,a} = \vec{R}) \cdot Pr_{\mathcal{H}}(H)$$

$$\leq (1 + \frac{1}{k}) \cdot Pr_{\mathcal{H}}(H)$$

and we are done with the right inequality of Lemma 11. The left inequality follows from the symmetric counterpart of Equation 6 and we are done with the proof of Lemma 11. □

**Proof of Lemma 9:** We start with the first part of Lemma 9. Let $\mathcal{R}'$ be an arbitrary set of measure at least .01 in $\mathcal{R}$ and suppose that if $R$ is randomly and uniformly chosen in $\mathcal{R}$ then the probability of getting a good vertex for the splice in Experiment 1 is at least $1/k^\gamma$ (as guaranteed by Lemma 8). We will show that the probability that $P_S$ succeeds with the verifier choosing its coin tosses $R \in \mathcal{R}'$ is at least $\frac{99}{100}\frac{1}{k^\gamma}$.

Recall that for a "normal" uniformly chosen $R \in \mathcal{R}$, we can check the probability that $P_S$ succeeds by choosing $H$ according to $\mathcal{H}$, choosing a random tape for the simulator uniformly, choosing a random $(i, a)$ with probability $\frac{cf(k-i+1)}{N}$ as above, and checking if $(i, a)$ is a good vertex in the proof tree of $\hat{V}_H$ and $S_s$. The difference now is that $R$ is chosen from the set $\mathcal{R}'$ and so in order to compute the probability that the simulation based prover $P_S$ succeeds, we must think of the following experiment. An address $(i, a)$ is chosen as above, a random tape for the simulator $s$ is chosen uniformly, $H$ is chosen according to $\mathcal{H}$, $j$ is chosen uniformly between 1 to $m$, $R$ is chosen uniformly in $\mathcal{R}'$, and then we run $P_S$ against $S_s$ with $\hat{V}_H$ with $R$ spliced into the vertex $(i, a)$ at index $j$. Alternatively, we may think of this probability as choosing uniformly independently at random $s$ and $(i, a)$, then choosing $H$ according to $\mathcal{H}(i, a, s, \mathcal{R}')$, and finally, checking whether $(i, a)$ is good in the interaction of $\hat{V}_H$ with $S_s$. Let $I(s, H, i, a)$ be an indicator variable indicating if the vertex $(i, a)$ in the interaction between $S_s$ and $\hat{V}_H$ is good. Using this notation with $R$ being randomly chosen from $\mathcal{R}$ as in Lemma 8 the probability $p$ that the simulation based prover succeeds satisfies:

$$p = \sum_{i=1}^{k}\sum_{a=1}^{N} \frac{cf(k-i+1)}{N} \sum_{s \in \{0,1\}^{|s|}} \frac{1}{2^{|s|}} \sum_{H} Pr_{\mathcal{H}}(H)I(s,H,i,a) \geq \frac{1}{k^\gamma}. \tag{7}$$

Now, to switch to $R$ being chosen in $\mathcal{R}'$ we rewrite the probability of choosing $H$ in the above formula. The probability, $p'$, that the simulation based prover succeeds may be written as:

$$p' = \sum_{i=1}^{k}\sum_{a=1}^{N} \frac{cf(k-i+1)}{N} \sum_{s \in \{0,1\}^{|s|}} \frac{1}{2^{|s|}} \sum_{H} Pr_{\mathcal{H}(i,a,s,\mathcal{R}')}(H)I(s,H,i,a) \tag{8}$$

$$\geq \sum_{i=1}^{k}\sum_{a=1}^{N} \frac{cf(k-i+1)}{N} \sum_{(s,H) \notin \mathcal{C}} \frac{1}{2^{|s|}} \cdot Pr_{\mathcal{H}(i,a,s,\mathcal{R}')}(H) \cdot I(s,H,i,a)$$

We only have pairs $(s, H)$ that are not in $\mathcal{C}$ in the summation, and thus, we may apply Lemma 11 to relate the probability of picking $H$ under $\mathcal{H}(i, a, s, \mathcal{R}')$ to the probability of picking $H$ under $\mathcal{H}$.

$$p' \geq (1 - \frac{1}{k}) \cdot \sum_{i=1}^{k} \sum_{a=1}^{N} \frac{cf(k - i + 1)}{N} \sum_{(s,H) \notin \mathcal{C}} \frac{1}{2^{|s|}} \cdot Pr_{\mathcal{H}}(H) \cdot I(s, H, i, a)$$

Now, since the measure of all $(s, H) \in \mathcal{C}$ is negligible, we may add back $\mathcal{C}$ to the summation and subtract some negligible fraction $\epsilon$:

$$p' \geq (1 - \frac{1}{k}) \cdot \sum_{i=1}^{k} \sum_{a=1}^{N} \frac{cf(k - i + 1)}{N} \sum_{(s,H)} \frac{1}{2^{|s|}} \cdot Pr_{\mathcal{H}}(H) \cdot I(s, H, i, a) \ - \ \epsilon$$

$$= (1 - \frac{1}{k}) \cdot p - \epsilon \geq \frac{99}{100} \frac{1}{k^{\gamma}}$$

and we are done with the first part of Lemma 9. To show that the second part holds as well, we need to show that the upper bound on the bad vertices is also close to the one given in Lemma 8. To show this, we define $p_{bad}$ instead of $p$ in Equation 7 to represent the probability that we get a bad vertex when $R$ is chosen uniformly. We then define $p'_{bad}$, to represent the probability that we get a bad vertex when $R$ is chosen in $\mathcal{R}'$, instead of $p'$ in Equation 8. In order to remove pairs $(s, H) \notin \mathcal{C}$ from the sum while giving an upper bound, we compensate on the removal by adding a negligible fraction $\epsilon$. Here, we rely on the fact that $\mathcal{C}$ is negligible also when $H$ is chosen according to $\mathcal{H}(i, a, s, \mathcal{R}')$. Using Lemma 11 again to relate the probability of picking $H$ under $\mathcal{H}(i, a, s, \mathcal{R}')$ to the probability of picking $H$ under $\mathcal{H}$. We get in the end that $p'_{bad}$ is at most $(1 + \frac{1}{k}) \cdot p + 2\epsilon$, and so $p'_{bad} \leq \frac{101}{100} \cdot p_{bad}$ and we are done with the proof of Lemma 9. □

## 6 Acknowledgments

We thanks Amit Sahai for useful comments.